\documentstyle[psfig]{mn}
\def\H0{{\it H}$_0$}

\def\q0{{\it q}$_0$}

\def\ergps{erg~s$^{-1}$}

\def\nH{$N_{\rm H}$} 
\def\psqcm{cm$^{-2}$}
\def\ergpspsqcm{erg~cm$^{-2}$~s$^{-1}$}

\def\phpspsqcm{ph\thinspace s$^{-1}$\thinspace cm$^{-2}$}
\def\phpkeVpspsqcm{ph\thinspace keV$^{-1}$\thinspace s$^{-1}$\thinspace cm$^{-2}$}
\def\ltsima{$\; \buildrel < \over \sim \;$}
\def\simlt{\lower.5ex\hbox{\ltsima}}
\def\gtsima{$\; \buildrel > \over \sim \;$}
\def\simgt{\lower.5ex\hbox{\gtsima}}

\title[ASCA/ROSAT observations of NGC5548]
{ASCA and ROSAT observations of NGC5548: discrepant spectral indices}
\author[K. Iwasawa, A.C. Fabian \& K. Nandra]
{\parbox[]{6.5in}{K. Iwasawa$^1$, A.C. Fabian$^1$ and K. Nandra$^2$}\\
\\
$^1$ Institute of Astronomy, Madingley Road, Cambridge CB3 0HA\\
$^2$ NASA/Goddard Space Flight Center, Greenbelt, MD 20771, USA} 

\date{}

\begin{document}

\maketitle

\begin{abstract}
We report on simultaneous ASCA and ROSAT observations of the Seyfert
galaxy NGC5548 made during the ASCA Performance Verification phase.
Spectral features due to a warm absorber and reflection are clearly
seen in the X-ray spectra. We find that the continuum spectral shape
differs between the ASCA and ROSAT datasets. The photon-index obtained
from the ROSAT PSPC exceeds that from the ASCA SIS
$\Delta\Gamma\approx 0.4$. The discrepancy is clear even in the 0.5--2
keV energy band over which both detectors are sensitive. The spectra
cannot be made consistent by choosing a more complex model.
The problem likely lies in the response curve (estimated effective
area) of one, or both, detectors. There may be important consequences
for a wide range of published results.
\end{abstract}


\section{Introduction}

The soft X-ray band below 2~keV has been covered by all recent X-ray
imaging satellites. The energy band is rich in spectral features
(absorption, emission lines, excess soft X-ray emission and so on),
which are useful diagnostic tools for understanding the nature of an
X-ray source of interest. It is therefore important that the spectra
are clearly understood and that data from different satellites are
calibrated reasonably well with each other so that results are
independent of the instrument used.

It has been noticed that there are sometimes significant discrepancies
between spectral measurements taken with ROSAT and ASCA (e.g., Yaqoob
et al 1994; Allen \& Fabian 1997; Ptak 1997). Typically, for AGN,
steeper spectral slopes are obtained from ROSAT PSPC measurements than
from ASCA. The steep ROSAT spectral slopes have sometimes been
interpreted as due to excess soft X-ray emission, when compared with
measurements from hard X-ray experiments (e.g., EXOSAT ME, Ginga LAC,
and ASCA). However, similar discrepancies have been found in quasar
spectra between the PSPC and the Einstein Observatory IPC (Fiore et al
1994; Laor et al 1994; Ciliegi \& Maccacaro 1996). As the bandpass of
the two detectors largely overlaps, calibration errors have been
suspected (Fiore et al 1994).

We aim in this paper to address the present status of the discrepancy
between the ASCA SIS and ROSAT PSPC, using data from the bright
Seyfert galaxy NGC5548 for which there was simultaneous coverage with
both satellites. Although the soft X-ray spectrum of NGC5548 is not
ideal for calibration because of complexity due to a warm absorber and
excess soft X-ray emission, the high spectral resolution of the ASCA
SIS is capable of resolving these features. Since the observation was
carried out before any significant degradation of the SIS detector had
occurred, the SIS spectrum has good spectral resolution. There are
several other simultaneous ASCA/ROSAT observations of AGN, but strong
absorption or more complex soft X-ray spectra of these targets (e.g.,
IRAS18325--5926, Iwasawa et al 1996) make cross-calibration difficult.
C. Otani (priv. comm.) finds a similar discrepancy from the
simultaneous ASCA/ROSAT observations of MR2251--178 to that reported
here. NGC5548 is stable and bright in the soft X-ray band where both
the ASCA SIS and ROSAT PSPC are sensitive; it has also been well
studied (Nandra et al 1993; Done et al 1995; Reynolds 1997; George et
al 1998). Preliminary results from efforts of cross-calibration
between ASCA and other satellites, e.g., BeppoSAX and RXTE, are now
available and the consistency between them is discussed.

In Section 2, we give detailed information on the data reduction. In
Section 3, we describe the integrated ASCA spectrum from the whole
observation. The ROSAT PSPC data are presented in Section 4. The
simultaneous data are described in Section 5, where we also examine
the consistency of the two datasets. The results and their implication
are discussed in Section 6.

\section{Observations and data reduction}


\begin{table*}
\begin{center}
\caption{FTOOLS and calibration files used for the ASCA SIS data
reduction and analysis. $^{\ast}$Charge Transfer Inefficiency (CTI) is caused
by radiation damage of the CCDs and has been increasing as a function 
of time.}
\begin{tabular}{lll}
Task & FTOOLS & Calibration files \\[5pt]
PI-channel conversion & SISPI (version 1.1) & sisph2pi$\_$110397.fits 
(CTI$^{\ast}$ table) \\
Response matrix & SISRMG (version 1.10) & \\
Detector efficiency & ASCAARF (version 2.72) & ${\rm xrt\_ea\_v2\_0.fits}$
(XRT effective area) \\
& (Effective area fudge ON, & ${\rm xrt\_psf\_v2\_0.fits}$ (XRT point spread function) \\
& Arf filter ON) & \\
\end{tabular}
\end{center}
\end{table*}

The ASCA observation started at 1993 July 27, 15:36 (UT) and ended at
1993 July 28, 08:24.
The Solid state Imaging Spectrometer (SIS; S0 and S1) was operated
using 4 CCD chips switching between Faint and Bright modes with 
High and Medium telemetry rates, respectively.
No level discriminater was applied.

The ASCA data reduction was performed using FTOOLS (version 4.0) 
and standard calibration files
provided by the ASCA Guest Observer Facility (GOF) at NASA/Goddard Space
Flight Center.
We used only the SIS data here. 
The Faint mode data were converted to the format of Bright mode so that
all the data could be added together. 
This means that the correction for the `Echo' and `Dark Frame Error' (DFE) 
(Otani \& Dotani 1995) cannot be applied. 
Since the observation was carried out only five
month after the launch of the satellite, those effects on the
data are small and taken into account in the detector response
matrices generated by SISRMG (version 1.1). 
Another effect known to degrade the SIS data is Residual Darkframe
Distribution 
(RDD, Dotani et al 1998), which is dark current remained after the DFE
correction due to non-uniform dark level between CCD pixels, 
although it should be small for this early ASCA observation as
the accumulation of the radiation damage is the main cause.
We have verified that the Faint mode data, taken during this observation,
with DFE and Echo corrections applied
give entirely consistent results with the same data in Bright mode,
apart from the normalization.
The Bright mode data give $\sim 5$ per cent smaller normalization than
the Faint mode data, which can be attributed to the RDD effect.
The RDD effect is largest when all 4 CCD chips are operating,
as in the present observation.
It causes a reduction of efficiency which is almost
independent of energy. The reduction of efficiency 
by $\sim 5$ per cent is in the range expected from the RDD effect.
The SIS results shown below are not corrected for this 
efficiency reduction.

Observed events with grades of 0, 2, 3 and 4
were selected and hot/flickering pixels on the CCD chips were removed.
The data selection criteria are, 1) source elevation higher than 5$^{\circ}$
and 25$^{\circ}$ above the night and bright Earth rims, respectively; 
2) cut-off rigidity greater than 4 GeV c$^{-1}$; and
3) X-ray telescope pointing fluctuation is less than 30 arcsec.
Data taken within 120 seconds of the spacecraft passing through
the transition from bright to night Earth, and the
South Atlantic Anomaly, are discarded.
The total good exposure time for each SIS detector was 30 ks.
The details of the calibration files we used are given in Table 1.

The source data were collected from a circular region with a radius of 
4 arcmin (see the details in Table 2) 
while the background data were taken from a source-free region
in the same field of view on the detector. 
The background counts in the 0.5--10 keV band are only 1 per cent 
of the source counts in both SIS spectra.
The effective area as a function
of energy, appropriate for the spectra obtained, was computed with 
ASCAARF (version 2.72), assuming a point source lying at the centre of the
photon collecting region (`point'=yes, `simple'=yes in ASCAARF, see also 
Table 1).

Note that the source extraction regions spread over the four CCD chips
in each detector, although most of the photons are collected on the
main chip (S0 chip-1, S1 chip-3). Therefore some photons fell into the
interchip gaps. The effective area of the ASCA X-ray telescope (XRT)
is believed to be reasonably accurate for a point source when
integrated over the image, but the azimuthal dependence of the point
spread function (PSF) may not be modelled accurately, as its shape is
like a `maltese cross'. We have made an experiment to check this
issue, using various source extraction-regions with and without
interchip gaps in them. The effective area of each spectrum was
computed with ASCAARF. A comparison of the fluxes obtained from those
spectra shows differences between of 2--3 per cent. Although ASCAARF
does not correct for photons falling in the interchip gaps, the
resulting error in the flux estimate is insignificant. A larger error
could result when the source centroid lies closer to these gaps.

During the ASCA observation, the ROSAT PSPC also observed the galaxy
for a short period, as shown in Fig. 1. 
The ROSAT coverage was from 1993 July 28, 00:53 to 09:16.
Total exposure time of the PSPC data is 4.3 ks of which about 3 ks was
covered also by ASCA.
We used the ROSAT data with the standard SASS processing.
The PSPC spectrum of NGC5548 was reduced using XSELECT.
The source photons were collected from a circular region with a radius of
1.9 arcmin while the background data are taken from an anulus of 4.0--5.8
arcmin radii centred on the source.
The response matrix for the PSPC data (`pspcb$\_$gain2$\_$256.rsp'), 
which is appropriate for 
a point source located at the centre of the field of view, is taken from 
the ROSAT calibration database.
We have tested the data corrected for temporal and spatial 
gain variations in the PSPC
by PCPICOR (which also fixes a bug in the SASS processed data) 
and found no significant difference from the original
data for which results are shown here.


\begin{table}
\begin{center}
\caption{The source extraction regions for the ASCA SIS (S0 and S1).
They are circular regions of $\approx 4$ arcmin radius
centred on each peak of the X-ray images.
The centre and radius are shown in unit of pixel in the detector
coordinates (DETX, DETY).}
\begin{tabular}{ccc}
Detector & Centroid & Radius \\[5pt]
S0 & (619,\thinspace 555) & 147 \\
S1 & (615,\thinspace 599) & 147 \\
\end{tabular}
\end{center}
\end{table}


\begin{figure}
\centerline{\psfig{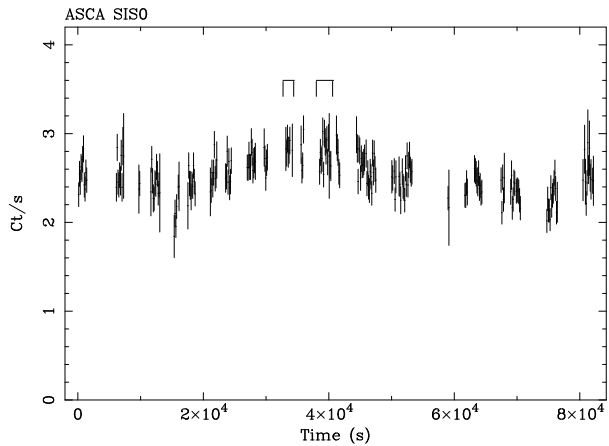}}
\caption{The 0.4--10 keV light curve of NGC5548 observed with the
ASCA SIS0. The epoch of the light curve is 1993 July 27, 15:36 (UT).
Each bin has an exposure of 128 s. The ROSAT PSPC simultaneously observed
the periods (3.34--3.42)$\times 10^4$ s and (3.84--4.02)$\times 10^4$ s,
which are indicated in the light curve.}
\end{figure}

\section{The ASCA data}


\begin{figure}
\centerline{\psfig{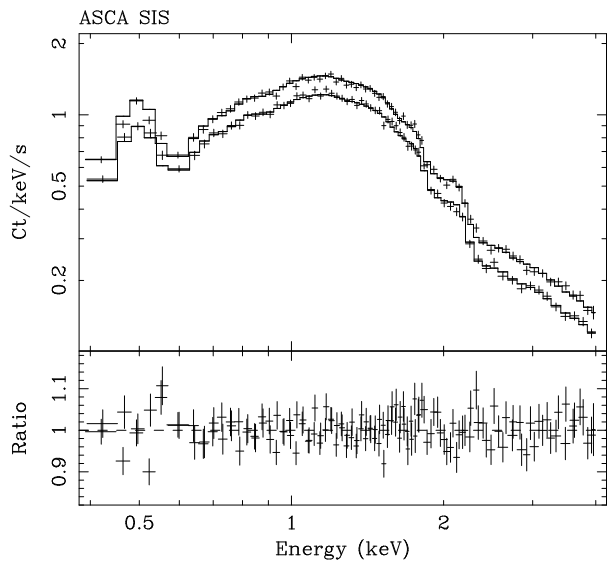}}
\caption{The soft X-ray part of the integrated ASCA SIS spectrum of NGC5548.
The data from the S0
and S1 are plotted with the best-fit model of a power-law plus three
absorption edges (see Table 3).}
\end{figure}

The ASCA 0.4--10 keV light curve shows moderate flux variation
across the observing run (Fig. 1). We first describe the spectral properties 
of NGC5548 in the ASCA band (see also Reynolds 1997; George et al 1998), 
using the integrated whole ASCA SIS dataset and then
discuss the data simultaneously observed with both satellites.
Spectral analysis was performed using XSPEC (version 10.0) and the
data from the two SIS detectors are fitted jointly.

A simple power-law fit to the whole band data leaves significant residuals 
(see Fig. 1 in Reynolds 1997; note that the upturn of the residual below
1 keV, e.g., in the EXOSAT ME/LE spectrum, has sometimes been interpreted 
as a soft excess, but it turns out to be an effect of a warm absorber). 
NGC5548 has been known to show spectral 
features typical of Seyfert 1 galaxies (e.g., Mushotzky, Done \&
Pounds 1993) such as a warm absorber (Nandra et al 1991), 
iron K line and spectral hardening at high energies due to reflection 
from optically thick, cold matter (Nandra \& Pounds 1994). 
These spectral features modify most of the primary power-law except for 
a narrow energy range around 3 keV. We first investigate 
the soft X-ray spectrum, which is not affected by reflection.

We fit the 0.4--4 keV data to investigate spectral features due to the
warm absorber (Fig. 2).
At least three absorption edges are significantly detected (Table 3)
when a simple power-law is employed for the continuum. The photon-index of the
power-law is $\Gamma = 1.934\pm 0.011$ and no excess absorption
above the Galactic value (\nH\ $= 1.7\times 10^{20}$\psqcm, 
Dickey \& Lockman 1990) is required. Apart from 
a weak line-like feature at 0.55 keV, most likely due to an 
under-subtracted atmospheric oxygen line, the quality of the fit is 
excellent ($\chi^2 = 260.44$ for 238 degrees of freedom, see Fig. 2). 
The three absorption edges are identified with OVII, OVIII and NeIX.
There is no obvious signature of extra soft X-ray emission.

The presence of a weak soft excess has been reported by Reynolds
(1997) using the same data.  However, the model he used for the fit
consists of only the two oxygen edges and a simple power-law for the
0.6--10 keV data despite the presence of the 1.2 keV edge and spectral
hardening above 4 keV due to reflection presented here.  Although a
blackbody-type soft excess is indeed significantly detected when only
two edges are considered in the model for the 0.4--4 keV data, the
inclusion of the 1.2 keV edge instead provides a better fit to the 
data.


\begin{figure}
\centerline{\psfig{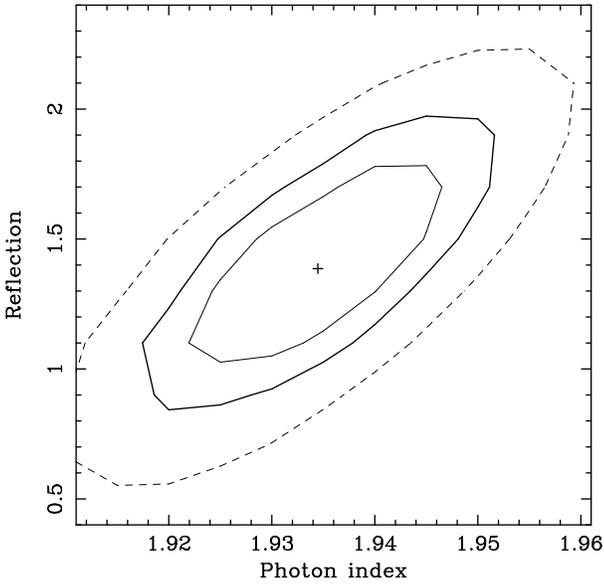}}
\caption{Confidence contours between photon-index and reflection strength
when the integrated ASCA 0.4--10 keV data are fitted with the model given 
in Table 3.
The 68, 90 and 99 per cent confidence levels for two interesting parameters
are plotted.}
\end{figure}

An extrapolation of the best-fit model shows that 
the continuum slope becomes flatter above $\sim$4 keV towards 10 keV
(see Fig. 4).
The iron K line is clearly detected (see also Mushotzky et al 1995).
A simple power-law fit to the 4--10 keV band excluding the Fe K band 
(5.5--7 keV) gives a photon-index $\Gamma = 1.82^{+0.10}_{-0.09}$.
An absorption edge feature around 8 keV, observed in the Ginga spectra 
(Nandra \& Pounds 1994), is 
barely detected at $8.1\pm 0.4$ keV ($\tau =0.15^{+0.18}_{-0.13}$).
A likely origin for this is K-shell absorption of partially ionized iron
in the warm absorber. When the iron K edge is included in the fit,
a slightly flatter power-law slope, $\Gamma = 1.76\pm 0.12$, is inferred.
The spectral hardening above
10 keV has been observed clearly in the Ginga spectra (Nandra \& Pounds 1994).
X-ray reflection from cold matter can account for this hard tail
(e.g., George \& Fabian 1991). 


\begin{figure}
\centerline{\psfig{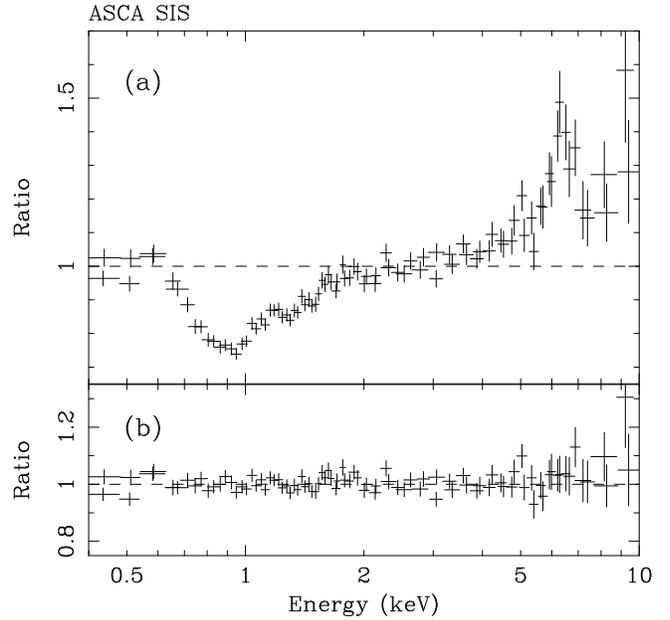}}
\caption{(a) Spectral features observed in the integrated 
ASCA SIS spectrum. Deviations
of the data from the power-law ($\Gamma = 1.934$) modified by the
Galactic absorption are plotted, removing the warm absorption, reflection
and iron line features from the best fit model which gives residuals
shown in lower panel (b). Note that there are alternative models to describe
the data (see text).}
\end{figure}


\begin{table*}
\begin{center}
\caption{Spectral fits to the ASCA data. The warm absorption is described
with three absorption edges. The threshold energies for absorption
edge and line energies are given in the source rest frame. The power-law
reflection model, pexrav (Magdziarz \& Zdziarski 1995), in XSPEC is used with
the Galactic absorption.  
$^{\ast}A$ is the normalization of power-law in units of \phpkeVpspsqcm\ 
at 1 keV. $^{\dag}$The value for the cosine of inclination of 
the reflecting slab, $i$, is fixed at 0.95 (or $i\sim 20^{\circ}$). Strength
of reflection, $I_{\rm Refl}$, is unity when the reflection matter covers
half of the sky. The Galactic absorption, \nH\ $=1.7\times 10^{20}$\psqcm, is
assumed as no excess absorption is detected.}
\begin{tabular}{cccc}
\multicolumn{4}{c}{Absorption edges}\\[5pt]
& $E_{\rm th}$ & $\tau$ & ID\\
& keV & & \\[5pt]
(1) & $0.73\pm 0.02$ & $0.30\pm 0.05$ & OVII \\
(2) & $0.86\pm 0.02$ & $0.22\pm 0.06$ & OVIII \\
(3) & $1.24\pm 0.04$ & $0.056\pm 0.021$ & NeIX \\[10pt]
\end{tabular}
\begin{tabular}{ccccc}
\multicolumn{5}{c}{Fe K line (double-gaussian)}\\[5pt]
& $E_{\rm line}$ & $\sigma$ & $I$ & $EW$ \\
& keV & eV & $10^{-5}$\phpspsqcm & eV \\[5pt]
(1) & $5.94\pm 0.10$ & $75^{+125}_{-75}$ & $2.22\pm 1.41$ & $41\pm 26$ \\
(2) & $6.44\pm 0.04$ & $32^{+80}_{-32}$ & $4.27\pm 1.45$ & $91\pm 31$ \\[10pt]
\end{tabular}
\begin{tabular}{ccccc}
\multicolumn{5}{c}{Power-law reflection model}\\[5pt]
$\Gamma$ & $A^{\ast}$ & cos\thinspace $i^{\dag}$ & $I_{\rm Refl}$ & \nH \\
 & $10^{-2}$ & & & $10^{20}$\psqcm \\[5pt]
$1.934\pm 0.014$ & 1.87 & 0.95 & $1.39\pm 0.47$ & 1.7 \\
\end{tabular}
\end{center}
\end{table*}

The use of the power-law reflection model ({\tt pexrav}, 
Magdziarz \& Zdziarski 1995)
instead of a simple power-law provides a good fit to the 0.4--10 keV
data if the iron K line is also modelled with a double-gaussian.
Among the parameters of the reflection model, the 
abundance of iron and the other
elements in the reflecting matter is assumed to be 1 solar. 
As the inclination angle, $i$, of the reflecting slab (or the accretion disk)
cannot be well constrained (when the strength of reflection is assumed to be 
1.0, the upper limit is 80$^{\circ}$), cos $i = 0.95$ (or 
$i\simeq 20^{\circ}$) is assumed. The best-fit values for photon index and 
reflection strength are $\Gamma = 1.93\pm 0.02$ and 
$I_{\rm refl}=1.4\pm 0.5$, and 
a confidence contour plot for the two parameters is shown in Fig. 3.
The spectral features which have modified the primary power-law are 
demonstrated in Fig. 4a, and the model consisting of 
a power-law plus warm absorption, reflection and
an iron K line gives a good fit to the data ($\chi^2 = 453.09$ for
475 degrees of freedom, see Fig. 4b).
The data are thus consistent with the reflection from a (nearly) 
face-on accretion disk
illuminated by a power-law ($\Gamma\simeq 1.93$) source above it.

The observed fluxes are $2.7\times 10^{-11}$\ergpspsqcm\ in the 0.5--2
keV band and $4.4\times 10^{-11}$\ergpspsqcm\ in the 2--10 keV band.
The 0.5--2 keV flux corrected for the warm absorption is $3.1\times
10^{-11}$\ergpspsqcm. Note that the fluxes could be higher by $\sim
20$ per cent than the SIS values given above on account of the
comparison with GIS results.

The response matrix of the GIS (version 4.0) has a large systematic
error in the low energy range. The corresponding error in
\nH\ measurement is about $\Delta N_{\rm H}\simeq 1\times 10^{21}$\psqcm\
(private com., T. Dotani and the GIS team), implying that GIS spectra
below 1.5 keV are not as reliable as SIS spectra. The spectral
parameters obtained from the 2--10 keV GIS data are consistent with
the SIS results within errors, but with $\sim 20$ per cent larger
normalization than the SIS. Although about 5 per cent can be explained
by the RDD effect (see Section 2), the remaining $\sim 15$ per cent
may be an error in the flux estimated in the SIS. Telemetry
saturation, which is most severe in 4CCD observations during the PV
phase, might account for part of the error. The GIS has been flux
calibrated with the Crab Nebula, the classic standard X-ray source
used for flight calibration of many previous medium-hard X-ray
instruments. The source is too bright for the SIS to observe because
of pulse pile-up. For X-ray sources with a non Crab-like
spectrum, $\sim$5--10 per cent difference in flux between the GIS and
SIS are often seen, according to the calibration status report
published by the ASCA GOF.

The strength of reflection may be too strong. An acceptable fit to the data 
can be obtained by alternative models involving
no strong reflection below.
1) A single power-law
modified by warm absorption with deep edges in the 2--3 keV ranges which could
be attributed to K-edges of ionized Si, S and Ar, in addition to the
three edges considered above. The photon index would be $\Gamma\simeq 1.90$.
This model is far inconsistent with a one-zone
solar abundance warm absorber. 2) A double power-law discussed in George 
et al (1998). 
This model makes a curving continuum which steepens at lower energies.
There are two solutions for the combination of power-laws 
[PL$_i(\Gamma_i, n_i$), where $\Gamma$ and $n$ are photon-index
and normalization at 1 keV of a power-law and $i = 1, 2$] 
which provide similar 
quality of fit when the three absorption edges are included: 
a) $\Gamma_1\simeq 2.3$, $\Gamma_2\simeq 1.8$, $n_1/n_2\sim 0.4$; 
and b) $\Gamma_1\simeq 2.0, \Gamma_2\simeq 1.6, n_1/n_2\sim 3.6$.


\begin{figure}
\centerline{\psfig{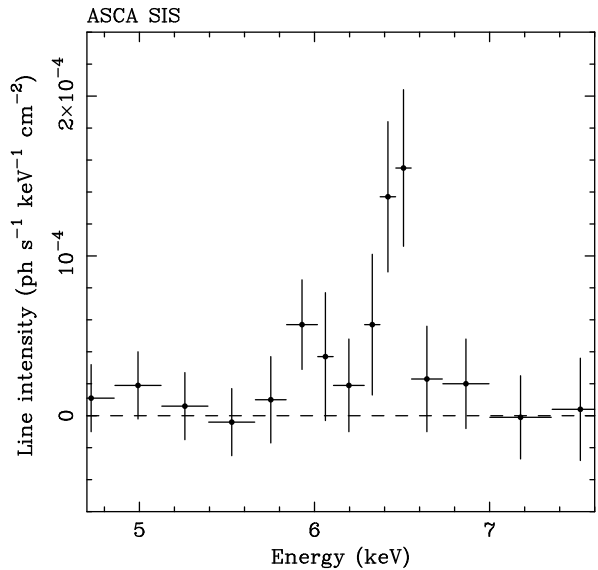}}
\caption{The iron K line profile of NGC5548 obtained from the integrated
ASCA SIS data. The two
SIS detectors are summed together. The energy scale is corrected for 
the galaxy redshift ($z=0.017$).}
\end{figure}

The shape of the iron K line was reported to be broad ($\sigma\sim 0.5$ keV)
by Mushotzky et al (1995). With the present analysis, the line profile
appears to be split
into two components: a major one at an energy of 6.4 keV and a weaker
one at 5.9 keV (Fig. 5 and Table 3, also see Otani 1995; Nandra et al 1997). 
The shape is difficult to explain with the ordinary diskline models
(e.g., Fabian et al 1989; Laor 1991). It may be the sum of a diskline and 
a narrow line at 6.4 keV.
The total equivalent width of the line is $\sim 130$ eV.

The photon-index of the primary power-law source is in a good agreement
with that derived from the simple power-law plus three absorption edge
model for the 0.4--4 keV data (Table 3).
This justifies the use of the ASCA data below 4 keV in investigating
the soft X-ray spectrum, obtained from the simultaneous observation 
presented below, without being affected by the reflection component.
It should however be noted that the primary power-law above 4 keV can
be flatter than this. The reflection model fit to the 2--18 keV Ginga 
data derived $\Gamma =1.81\pm 0.02$ and $I_{\rm Refl}=0.7\pm 0.4$ 
(Nandra \& Pounds 1994) with use of the Lightman \& White (1988) type 
reflection model ({\tt plrefl} in XSPEC). 
This can also describe the 4--10 keV ASCA data
when an iron K edge is included at around 8 keV.

\section{The ROSAT data}


\begin{table*}
\begin{center}
\caption{Spectral fits to the ROSAT PSPC data. Spectral models are 
(1) a simple power-law; and (2) a power-law modified by Galactic
absorption (\nH\ $= 1.7\times 10^{20}$\psqcm) and an absorption edge.
The threshold energy of the absorption edge is measured in the 
galaxy rest frame.}
\begin{tabular}{ccccccc}
Model & $\Gamma$ & $A$ & \nH & $E_{\rm th}$ & $\tau$ & $\chi^2$/dof \\
&& $10^{-2}$ & $10^{20}$\psqcm & keV & & \\[5pt]
(1) & $2.37\pm 0.04$ & 1.74 & $1.6\pm 0.09$ & --- & --- & 191.52/176 \\
(2) & $2.35\pm 0.02$ & 1.87 & 1.7 & $0.80\pm 0.06$ & $0.23\pm 0.10$ & 177.34/175 \\
\end{tabular}
\end{center}
\end{table*}


\begin{figure}
\centerline{\psfig{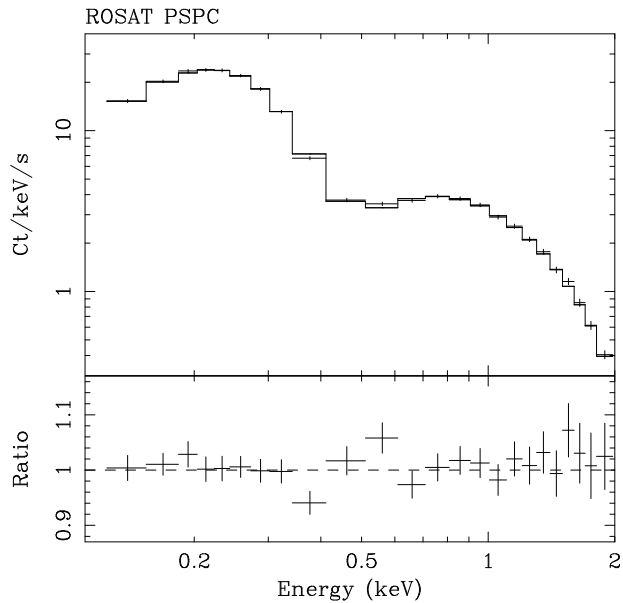}}
\caption{The ROSAT PSPC spectrum of NGC5548 fitted with a 
power-law plus an absorption edge model (Model (2) in Table 4).}
\end{figure}

We now present the spectral analysis of the total (4 ks) ROSAT PSPC data.
A simple power-law fit to the 0.14--2 keV data 
gives photon index $\Gamma = 2.38\pm 0.04$
and absorption column density of $(1.65\pm 0.09)\times 10^{20}$\psqcm,
which is consistent with the Galactic value. 
The inclusion of an absorption edge
improves the quality of the fit significantly (Table 4). 
It is clear that the PSPC spectrum gives a steeper photon index 
($\Gamma = 2.35\pm 0.02$) than
the ASCA SIS spectrum.
The edge energy $\sim 0.8$ keV is an intermediate value of those of the
two oxygen edges (OVII and OVIII, see Table 3). 
Little improvement ($\Delta\chi^2 = 0.02$) 
is seen when a further edge is added.
The PSPC data and the best-fit model (2) are shown in Fig. 6.
There is no signature of excess soft emission.

The observed fluxes are $6.4\times 10^{-11}$\ergpspsqcm\ in the 0.1--2 keV
band and $3.7\times 10^{-11}$\ergpspsqcm\ in the 0.5--2 keV band,
estimated from the best-fit power-law plus edge model.
The 0.5--2 keV flux corrected for the absorption edge is 
$3.9\times 10^{-11}$\ergpspsqcm.

NGC5548 is known to show soft X-ray flares seen below 0.4 keV
(Done et al. 1995). The flux change of the soft X-ray component is 
not correlated with the higher energy continuum (Done et al 1995; 
Kaastra et al 1998). Here we investigate whether the source is
undergoing a similar soft flare. This is important when excess soft
X-ray emission is disscussed because this 
component appears out of the ASCA bandpass.

A monitoring campaign of NGC5548 with ROSAT PSPC was carried out between
1992 December and January 1993, 6--7 months before the present observation
(Done et al 1995).
During the campaign, a large flare over 8 days was observed with strong
spectral softening. 
Apart from the flare, the rest of the data show a 
similar spectral shape, judging from the fairly constant count rate ratio 
in the energy bands of 1.0--2.5 keV and 0.1--0.4 keV.
We thus compared our PSPC data with the pre-flare data 
(the dataset {\tt 01b} in Done et al 1995), 
which Done et al (1995) used as a template spectrum
and probably a representative one in normal states.
The simple power-law fit to the present data (Table 4) gives very similar
parameters to those for the {\tt 01b} dataset ($\Gamma = 2.36\pm 0.01$ and 
$A = 1.69\times 10^{-2}$\phpkeVpspsqcm, Done et al 1995);
the difference in normalization is only $\sim$2 per cent.
The value of the hardness ratio (0.32) for the present observation is 
also similar to that for the {\tt 01b} dataset.
Therefore we conclude that the X-ray source was in its normal state when the
simultaneous observations were carried out.

\section{The ASCA/ROSAT simultaneous data}


\begin{table*}
\begin{center}
\caption{Spectral fit to the ASCA data for the simultaneous coverage.
The 0.4--4 keV SIS data are used. The Galactic value \nH\ = 
$1.7\times 10^{20}$\psqcm\ is assumed for absorption of the power-law.
Two absorption edges are significantly detected. They are attributed to
partially ionized oxygen (OVII and OVIII) in the warm absorber. The
threshold energy of the absorption edges is 
measured in the galaxy rest frame.}
\begin{tabular}{ccccccc}
$\Gamma$ & $A$ & $E_{\rm th1}$ & $\tau_1$ & $E_{\rm th2}$ & $\tau_2$ & 
$\chi^2$/dof \\
& $10^{-2}$ & keV & & keV & & \\[5pt]
$1.954\pm 0.038$ & 1.83 & $0.74^{+0.05}_{-0.07}$ & $0.30^{+0.15}_{-0.16}$ &
$0.87^{+0.05}_{-0.04}$ & $0.27^{+0.15}_{-0.16}$ & 174.83/182 \\
\end{tabular}
\end{center}
\end{table*}


\begin{figure}
\centerline{\psfig{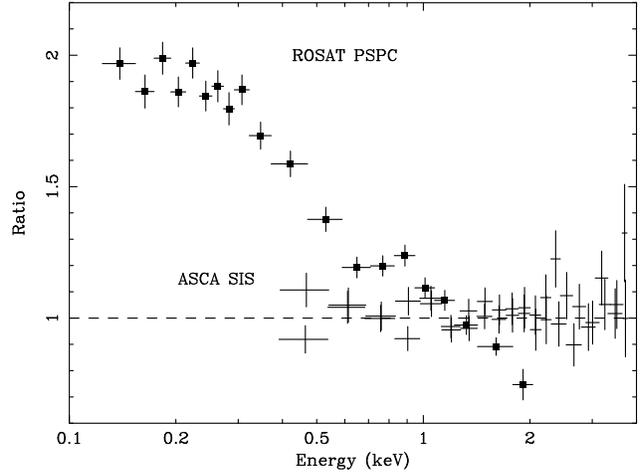}}
\caption{Ratio of the data of the ROSAT PSPC (filled squares)
and the ASCA SIS (crosses) to the best-fit model for the ASCA 0.4--4
keV spectrum are shown. The datasets are obtained from the
simultaneous observation. The SIS normalization has been adjusted to
agree with that of the GIS. A discrepancy in the spectrum between the
two instruments is clearly seen, even in the energy band covered by
both detectors (e.g., 0.4--2 keV band).}
\end{figure}

The period over which both ASCA and ROSAT observed NGC5548 were simultaneous
for about 3 ks. 
This is a large fraction of the total PSPC observation, during which 
no flux change is observed. The spectral data for the 3 ks is 
virtually identical to the total 4 ks spectrum.
On the other hand, the ASCA data of the simultaneous observation is 
a small fraction ($\sim 1/10$) of the total exposure.
The ASCA 0.5--2 keV flux during this period is 
$3.0\times 10^{-11}$\ergpspsqcm,
$\sim$ 10 per cent higher than the mean value for the whole observation.
The spectrum can be different from the averaged one if not dramatically.
We therefore analyze the ASCA spectrum and then compare it with 
the PSPC spectrum to illuminate any discrepancy in spectral measurement 
between the two satellites.

Since the spectrum below 4 keV is not affected by reflection, as 
discussed in Section 3, a simple power-law modified by
the Galactic absorption and a warm absorber was fitted to the 
ASCA SIS 0.4--4 keV data. Results of the spectral
fit are given in Table 5. 
The best-fit value of photon-index is slightly larger than that for
the whole dataset, but consistent within uncertainties.
Two absorption edges are significantly detected and are identified
with OVII and OVIII.

In comparing the ROSAT PSPC and the ASCA SIS, we plot the ratio
of the data from both detectors to the best-fit model to the 
ASCA 0.4--4 keV spectrum given in Table 5 (Fig. 7).
The SIS normalization has been adjusted to match
that of the GIS, being in favour of the absolute flux calibration in the GIS
(see Section 3).
The significantly better spectral resolution of the ASCA SIS compared to
the PSPC should provide a better description of the observed spectrum 
of the source. The deviation of the PSPC data from the ASCA data is clear.
If both detectors are well calibrated with each other, 
the PSPC data would lie around the ratio of unity, at least 
in the common 0.5--2 keV band.
A steep slope of the PSPC spectrum is however evident in the energy band.
The inclusion of any extra soft X-ray component below 0.4 keV (e.g., a
steep power-law or blackbody emission)
in addition to the ASCA best-fit model fails to provide an acceptable 
fit to the PSPC data ($\chi_{\nu}^2>4$).
This rules out the possibility that the excess seen in Fig. 7 can be due to the
energy response of the PSPC when there is a strong soft excess below 
the ASCA energy range. Even if the double power-law 
(see Section 3) is used for the continuum model, the diecrepancy in ratio
in Fig. 7 is reduced no more than 15 per cent.


The 0.5--2 keV fluxes for the simultaneous data 
from the two instruments are
$3.7\times 10^{-11}$\ergpspsqcm\ from the PSPC and 
$3.0\times 10^{-11}$\ergpspsqcm\ from the SIS. 
As mentioned in Section 3, the ASCA SIS flux 
could be
$\sim 20$ per cent larger than the quoted value, and 
thus becomes consistent with the ROSAT flux if the 
GIS normalization is trusted.

\section{DISCUSSION}

We now examine possible reasons for the discrepancy between ROSAT and
ASCA in spectral measurements. Steep ROSAT spectra have been noticed
in several analysis on AGN spectra by comparing Einstein Observatory
IPC (Fiore et al 1994; Laor et al 1994; Ciliege \& Maccacaro 1996),
EXOSAT ME (Shartel et al 1997), or BeppoSAX LECS (Mineo et al 1999)
measurements. As mentioned in the Introduction, a similar discrepancy
to the present result was reported from the ASCA/ROSAT
simultaneous observation of the quasar MR2251--178 (C. Otani, priv.
comm.), suggesting that the result presented in this paper is typical.
Since the ROSAT bandpass (0.1--2.4 keV) extends to lower energies than
the IPC ($\sim$0.5--4 keV), the steep spectra from ROSAT observations
have been interpreted as soft excess emission. This is a problem which
is particular to AGN spectra (however, it should be noted that the
PSPC data of quasars investigated by Laor et al (1994) are consistent
with a single power-law and no soft excess component below 0.5 keV is
required). For Seyfert galaxies like NGC5548, reflection also makes
spectral slopes flat in the ASCA band ($>4$ keV). These effects from
the difference in bandpass between satellites could result in
different measurements of spectral slope.

We found a difference of photon-index ($\Delta\Gamma\simeq 0.4$)
measured with the ROSAT PSPC
and the ASCA SIS during the simultaneous observation of NGC5548.
The ASCA data below 4 keV are not affected 
by reflection (see Section 3) in this object. 
Although we are not able to assess exactly how much the soft excess component
addes to the power-law continuum below 0.4 keV, we see no evidence for
a soft X-ray flare during the observation (see Section 4).
The ROSAT PSPC data are fully consistent with a simple power-law when
the warm absorption is taken into account (Fig. 5) and no extra soft
X-ray component is required.
As Fig. 7 demonstrates that the PSPC data are significantly
steeper than the ASCA data in the 0.5--2 keV band, we can rule out 
the possibility of bandpass effects for the different spectral
slope measurements. Therefore errors in calibration between the
two satellites are the likely explanation.

NGC5548 was observed with the ROSAT PSPC several times.  The first
observation was carried out during the ROSAT PV phase (1990 July
16--21) with a different detector (PSPC-C), which ceased operating on
1991 Jan 25; the results were published by Nandra et al (1993).
The observed flux suggests that the source during the observation was
in its low state with no soft X-ray flare taking place, but a weak
excess soft X-ray component was found below 0.5 keV (Nandra et al
1993). A simple power-law fit to the spectrum gives $\Gamma = 1.98\pm
0.04$, which is slightly steeper than the ASCA value, but similar.
However, when NGC5548 was observed with the PSPC-C once again during
the ROSAT All Sky Survey (Shartel et al 1997) at a flux level similar
to the present observation and {\tt 01b}, a steep photon-index, $\Gamma =
2.4\pm 0.1$, was found (Shartel et al 1997).  After the change of
detector (PSPC-B), the ROSAT spectra show similarly steeper slopes,
$\Gamma\approx 2.37$ for the normal state spectrum (Done et al 1995;
this work).  The flat power-law slope reported in Nandra et al (1993)
is exceptional possibly because the source was in its low state,
as other Seyfert 1 galaxies often show an anti-correlation between
X-ray spectral slope and flux (e.g., Mushotzky, Done, \& Pounds 1994).

The calibration of the ASCA SIS for the soft X-ray band has been
a controversial issue.
Photon-index and column density measurements of 3C58, a Galactic
synchrotron nebula, with Einstein Observatory IPC, EXOSAT ME/LE, Ginga LAC
and ASCA agree within uncertainties, although the scatter in \nH\ values 
derived from different CCD chips of the SIS imply a systematic error
(Dotani et al 1996).
General agreement on the systematic error in column density measurements 
indicates that the SIS 
over-estimates \nH\ value by 2$\times 10^{20}$\psqcm.
The other possible sources which could contribute to the systematic 
error of the SIS
calibration are a) DFE and Echo corrections in the response matrices
for the Bright-mode data used here; b) the RDD
effect; and c) uncertainty in the SIS gain.
The comparison with the Faint mode data has verified that the effects
of the DFE and Echo are taken into account in the response matrix
with enough accuracy. The RDD effect appears to reduce the efficiency 
by $\sim 5$ per cent, but it is unlikely to affect the 
spectral form, since the effect is almost energy independent.
The SIS gain is calibrated within 0.5 per cent accuracy (Dotani 
et al 1996). 
None of the possible systematic errors in the SIS data 
considered above is, therefore,
sufficient to explain the difference between the results from the 
two satellites in the observation of NGC5548 reported here.

Simultaneous observations of the quasar 3C273 with ASCA, RXTE and 
BeppoSAX have been recently carried out. 
The three satellites agree on the measurement of spectral slope 
($\Gamma\sim 1.62$ from ASCA SIS and GIS, RXTE/PCA, BeppoSAX/MECS). 
The BeppoSAX/LECS, which covers the energy range of 
0.1--4 keV, found an edge-like feature at 0.5 keV and a steep soft 
X-ray component below 0.4 keV (Grandi et al 1997). 
The SIS and the LECS are in a good agreement (Orr et al 1998).
Relative differences in \nH\ measured by the two detectors is 
reported, by Yaqoob et al (1997) using the latest calibration, as
$\Delta$\nH $<1.1\times 10^{20}$\psqcm\ for the S0 and 
$\Delta$\nH $<1.9\times 10^{20}$\psqcm\ for the S1.

During ASCA AO-5, NGC5548 was observed with ASCA and EUVE simultaneously.
The extrapolation of the ASCA spectrum matches the highest energy end
of the EUVE/SW measurement (Kaastra et al 1999; Kunieda et al 1998).
As the EUVE data cover the soft portion of the ROSAT bandpass,
this may provide further support for the ASCA result.

There are a number of discontinuities in the effective area curves of
the instruments originating from the materials used in the X-ray
telescope and detectors. These can cause a problem with the energy
calibration. In the ASCA response, uncertainties due to an oxygen edge
at $\sim$0.5 keV in the SIS detector and a gold edge at 2.2 keV in the
XRT (this residual is approximately modelled with ASCAARF) have been
noticed. A very deep carbon edge is present in the PSPC response,
which removes virtually all the efficiency in the 0.3--0.5 keV band.
Misplacement of an edge would introduce an excess or deficit of
effective area. This problem is more relevant to the PSPC than the
ASCA SIS, given the poor spectral resolution.

\section{Summary}

In summary, (1) the ROSAT PSPC yields a photon-index steeper than the ASCA
SIS by $\Delta\Gamma\simeq 0.4$ for the simultaneous observation of NGC5548.
(2) the steep spectral slope obtained from the PSPC cannot be accounted for by
any soft excess emission below the ASCA bandpass (below $\sim $0.5 keV).
(3) results on the spectral form 
from ASCA SIS data on other objects agree with 
that from other instruments; Einstein IPC, EXOSAT ME/LE, Gigna LAC, 
BeppoSAX LECS and MECS, and EUVE SW.
(4) the measured fluxes seem to be consistent between ASCA and ROSAT, when
the error of the SIS flux measurement is taken into account.

The discrepancy between ASCA and ROSAT also appears to affect 
measurements of {\it excess} absorption in clusters of galaxies. 
Allen \& Fabian (1997) reported that the excess absorption
column densities of the cooling flow cluster Abell 2199
obtained from the ASCA SIS and Einstein Observatory SSS agree well with
each other, but the ROSAT PSPC gives a significantly lower excess column
density.
The determination of excess column densities depends on the detailed
shape of the spectrum over the 0.5--1 keV range.
It is not clear, if Fig. 7 is a guide, that ROSAT PSPC results will
be correct in this respect.
If, however, the ROSAT PSPC energy calibration is correct, and the 
source of the discrepancy lies in its response amplitude, then
energy-dependent results such as total column density estimates
of all classes of source
and the temperatures of clusters where lines (e.g., the Fe-L complex)
are important will be unaffected by the problem discussed here.

Our results indicate that caution should be applied when (over)
interpreting differences between ROSAT and ASCA spectra. Given the
fact that the spectra in Fig. 7 depends on the exact spectral form of a
source, it is not advisable to try to correct other PSPC spectra using
this Figure, since it may lead to a false result. The predicted
spectral response curves of both satellites need reconsideration
(although an improved calibration of the PSPC has been attempted (S.
Snowden, priv. comm.), the calibration files are not publicly
available).

\section*{acknowledgements}

We thank all the members of the ASCA PV team and the ASCA Guest
Observer Facility at NASA's Goddard Space Flight Center. Tadayasu
Dotani is especially thanked for valueable comments on the SIS
calibration. 
Tahir Yaqoob and the referee are also thanked for their
helpful comments on the paper.  
We acknwledge the Royal Society (ACF),
PPARC (KI) and National Research Council of USA (KN) for support.

\bsp

\end{document}